\documentclass{article}
\usepackage{setspace}

\usepackage[english]{babel}

\usepackage[letterpaper,top=2cm,bottom=2cm,left=3cm,right=3cm,marginparwidth=1.75cm]{geometry}
\usepackage{caption}
\usepackage{amsmath}
\usepackage{graphicx}
\usepackage[colorlinks=true, allcolors=blue]{hyperref}
\usepackage{natbib}

\title{A Methodology to Quantify Interscale Energy Transfer at Solid Boundaries}
\author{Lennard Miller}
\date{}
\begin{document}
\maketitle

\begin{abstract}
    Far away from solid boundaries, energy can be transferred between different flow scales due to the non-linear self-advection of velocity. This energy transfer can be quantified using well-established Fourier diagnostics or filtering methods. However, these diagnostic tools fail to provide a physical representation of the linear energy transfer that may occur during the formation of oceanic boundary layers or Rossby wave reflections. In this document, I outline a novel filtering methodology that is able to quantify this linear energy transfer by combining coarse-graining with volume-penalization. Its utility is illustrated by quantifying the down-scale energy transfer occuring during a Rossby wave reflection off a western boundary. The conceptual framework developed here is thought to be broadly applicable to the study of multi-scale energetics of bounded geophysical fluid flows.
\end{abstract}

\section{Introduction}

The scales of oceanic flows range from millimeters for viscous flows to thousands of kilometers for synoptic phenomena. The study of these flows relies heavily on the idea that energy is conserved across this range of scales, meaning that in the absence of forcing and dissipation energy can be transferred between different flow scales but cannot be lost. A well-studied example of such interscale energy transfers are the inertial ranges of energy cascades in oceanic turbulence \citep{scott2005direct}. Non-linear self-advection of the flow velocity results in a flux of energy $\Pi$ across scales, which can readily be quantified using standard Fourier analysis or filtering methods \citep{sagaut2006large}. Interscale energy transfer can also occur due to linear processes at coastal boundaries, for example via the formation of dissipative western boundary layers \citep{stommel1948westward} or during the reflection of Rossby waves \citep{pedlosky2013geophysical}. However, Fourier analysis cannot capture the heterogeneous nature of these bounded flows and standard filtering approaches are ill-defined when the filter extends beyond the flow domain \citep{aluie2018mapping,hausmann2025formally}. Current spectral and filtering methods therefore fail entirely to quantify the linear interscale transfer of energy occurring at coastal oceanic boundaries.\\


In this document, I demonstrate that filtering methods can be extended to show that energy is conserved across scales in bounded flows. To this end, I adopt the framework of volume penalization \citep{engels2015numerical}. I begin by introducing an auxiliary porosity parameter $\chi$, which represents a “soft” boundary. I then construct an energy-conserving solution across the soft boundary that can be readily analyzed using standard filtering techniques. The case of a solid boundary is subsequently obtained in the limit $\chi \rightarrow 0$. This approach enables the definition of a novel interscale energy flux, $\Pi^\chi$, which quantifies the transfer of energy across scales at solid boundaries. As an illustration, I apply this methodology to compute the linear downscale energy flux which occurs when an oceanic Rossby wave reflects off a western continental boundary. The conceptual framework developed here is thought to be broadly applicable to the study of the multi-scale energetics of bounded geophysical fluid flows.\\

\section{Methodology: Combining Flow Filtering with Volume Penalization}
\label{filtering}
Filtering fluid flows is a classical method used to investigate the energy transfer between different flow scales \citep{sagaut2006large}. In this methodology, a filtered flow $\vec{u}_L$ is obtained via a convolution with a normalized filter $G_L$,
\begin{equation}
    \vec{u}_L(\vec{x}) = \int  G_L(\vec{x}'-\vec{x}) \vec{u}(\vec{x}') d\vec{x}'. \label{filter}
\end{equation}
The filtering kernel $G_L$ is a localized, normalized function whose width scales with $L$. The integral in \eqref{filter} is usually carried out across infinite fluid domains, or at least to a distance larger than $L$ beyond which the vanishing filter $G_L$ limits the contribution of the integrand to the filtered flow. However, it is unclear how this integral is to be carried out when the filter is moved closer than a distance $L$ to a solid boundary \citep{aluie2018mapping,hausmann2025formally}, as the flow outside of the domain is not a physically defined quantity. To circumvent this problematic, I here model solid boundaries following the volume penalization method as a sudden jump in porosity $\chi$. This method is well established as a numerical means to model fluid flows on domains with complex geometry \citep{schneider2005decaying, engels2015numerical}. Inside the domain the porosity is infinite and the fluid is allowed to move freely, but outside of the domain flow is restricted by a porous stress inversly proportional to $\chi$. The flow is then well defined both within and beyond the boundary, with the case of a solid boundary being achieved when $\chi\rightarrow 0$.\\

My aim in the following will be to quantify the downscale energy flux occuring during the reflection of oceanic Rossby waves. The simplest flow model that describes such a reflection is the linearized two-dimensional Navier-Stokes equation in a rotating frame of reference, given by
\begin{align}
    \partial_t\vec{u} + \vec{f}\times\vec{u} + \nabla P &=  - \frac{S(x)}{\chi}\vec{u} ,\label{NS_porous}\\
    \nabla.\vec{u} &= 0. \label{div}
\end{align}
Here $\vec{u} = (u,v)$ is the fluid velocity, $P$ is the fluid pressure and $\vec{f} = (f_0 + \beta y)\hat{z}$ is the background vorticity with $f_0$ the Coriolis frequency and $\beta$ the background vorticity gradient. The step function $S(x)$ is defined by 
\[
S(x) :=
\begin{cases}
1, & x < 0 \\
0, & 0 \le x 
\end{cases}
\]
and models an "soft" boundary at $x = 0$. I will now exemplify the derivation of a closed scale-dependent energy budget extending across a solid boundary using this dynamical framework.\\

An evolution equation of the filtered flow is readily obtained by applying the convolution \eqref{filter} to \eqref{NS_porous}. After integration by parts and noting that $\partial_{x}G_L = -\partial_{x'}G_L$ I obtain
\begin{equation}
    \partial_t\vec{u}_L + \vec{f}\times\vec{u}_L + \nabla P_L = \vec{\tau}^f_L + \vec{\tau}^\chi_L  \label{filtered^NS}
\end{equation}
The left hand side is identical to the unfiltered equation \eqref{NS_porous}, but there are now also the Coriolis subgrid stress $\vec{\tau}^f_L$ and, a novel quantity, the boundary subgrid stress $\vec{\tau}^\chi_L$. They are given by
\begin{align}
    \vec{\tau}^f_L &= \vec{f}\times\vec{u}_L  - (\vec{f}\times\vec{u})_L \\
    \vec{\tau}^\chi_L &=  \left(- \frac{S(x)}{\chi}\vec{u}\right)_L
\end{align}
We can then define the energy of the filtered flow $E_L$ as
\begin{equation}
    E_L = \frac{|\vec{u}_L|^2}{2}.\label{cumulative}
\end{equation}
An evolution equation for $E_L$ is then obtained by multiplying \eqref{filtered^NS} by $\vec{u}_L$. The resulting equation can be written as
\begin{equation}
    \partial_tE_L +\nabla.\left(\vec{u}_L P_L\right)= \Pi_L^f +\Pi_L^\chi,\label{Filtered_Energy}
\end{equation}
where $\Pi$ denote subgrid energy fluxes. Analogous to the subgrid stresses, there is the subgrid Coriolis Flux $\Pi_L^f$ and the subgrid boundary flux $\Pi_L^f$ defined by
\begin{equation}
    \Pi_L^f = \vec{u}_L.\vec{\tau}^f_L \quad , \quad \Pi_L^\chi = \vec{u}_L.\vec{\tau}^\chi_L
\end{equation}
Equation \eqref{Filtered_Energy} constitutes a closed energy budget for flow containing all scales larger than $L$. In the limit $L\rightarrow 0$, the energy equation of the unfiltered flow is recovered, implying that the term $\Pi_L^\chi$ vanishes. This means that $\Pi_L^\chi$ can transfer energy between scales but does not create nor dissipate energy as a whole. This demonstrates that the filtered energy $E_L$ is conserved across scales in bounded flows. We will now proceed to show that $\Pi_l^\chi$ quantifies the interscale energy transfer occurring at solid boundaries in the limit $\chi\rightarrow 0$.

\section{Proof of Concept: Downscale Energy Flux during Rossby Wave Reflections}

A typical example of downscale energy flux from geophysical fluid dynamics is the reflection of Rossby waves off a western boundary. An analytic solution that recalls the mechanics of such a reflection is derived in appendix \ref{appendix}, and will be only briefly summarized here.\\

\begin{figure}
    \centering
    \includegraphics[width=0.6\linewidth]{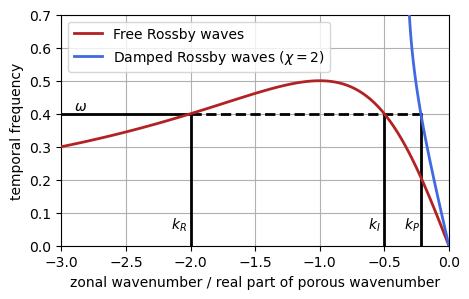}
    \caption{\textbf{Dispersion relation of free Rossby waves and the decaying branch of the modified dispersion relation in the porous medium.} The solution has been made non-dimensional by setting $\beta = f_0 = l = 1$. Upon reflection, an incoming wave of wavenumber $k_I$ excites a reflected wave of wavenumber $k_R$. When the incoming wave encounters a step in porosity an additional decaying porous wave with complex wavenumber $k_P$ is excited, whose real part is shown here for an exemplary value of $\chi = 2$.}
    \label{fig:Rossby_disp}
\end{figure}

\begin{figure}
    \centering
    \includegraphics[width=\linewidth]{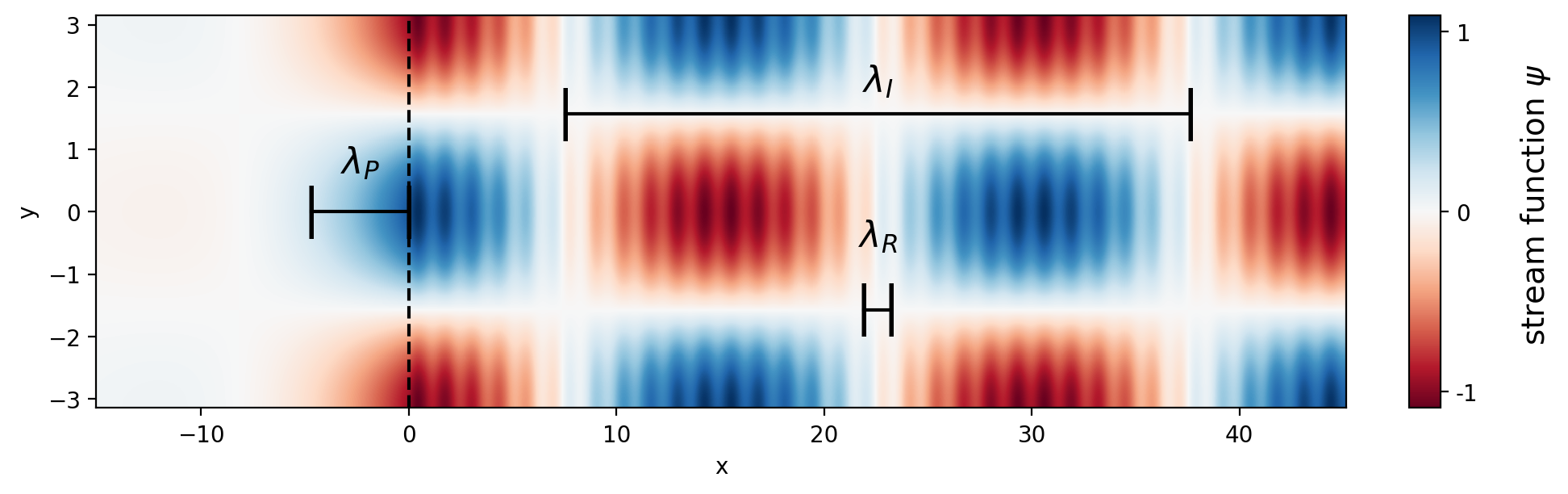}
    \caption{\textbf{Examplatory streamfunction for a Rossby wave reflection off a porosity step at $\mathbf{x = 0}$.} The solution shown here is derived in the appendix \ref{appendix}. The solution consists of an incoming wave of wavelength $\lambda_I$ which transports energy westwards. Upon encountering a step in porosity, this energy is partly reflected and carried eastward by a wave with $\lambda_R$, partly transmitted adn dissipated as the decaying wave over a scale $\lambda_P$. For this plot, the free parameters of the solution have been set to $\omega = 0.2$, $\chi = 5$, $A_I=0.5$, and the system has been made non-dimensional by setting $\beta = f_0 = l = 1$. }
    \label{psi_illustrate}
\end{figure}

A long, incoming Rossby wave with wavelength $\lambda_I = 2\pi/k_I$ carries energy towards western oceanic boundaries, where this energy is transferred downscale and carried eastward by a short, reflected wave of wavelength $\lambda_R = 2\pi/k_R$. These dynamics follow directly from the non-linearity of the dispersion relation of Rossby waves, shown in figure \ref{fig:Rossby_disp}. The reflected wave has to match the temporal frequency $\omega$ and the meridional wavenumber $l$ of the incoming wave, and is thus constrained to much shorter wavelengths. When the reflection occurs off a step in porosity with a finite value of $\chi$, an additional wave is excited in the porous medium which decays over a distance $\lambda_P = 1/\mathcal{I}(k_P)$ where $\mathcal{I}$ denotes the imaginary part. Such a solution is illustrated in figure \ref{psi_illustrate}. In the limit $\chi \rightarrow 0$ the amplitude of this porous wave vanishes and the case of a perfect reflection off a solid boundary is recovered.\\

We will here apply the filtering methodology outlined in section \ref{filtering} to the analytical solution for Rossby wave reflection off a porosity step given in the appendix \ref{appendix}, which will allow us to explicitly calculate the subgrid boundary flux $\Pi_L^\chi$.  In particular, I am interested in the asymptotic behaviour of the energy budget of the long, incoming wave in the limit $\chi\rightarrow 0$. In the following, all the diagnostics presented here are made non-dimensional by setting $\beta = f_0 = l = 1$. The temporal frequency of the waves are set to $\omega = 0.02$, which allows for a sufficient scale separation between $\lambda_I$ and $\lambda_R$ to apply the filtering. To counter the redundancy of the meridional and temporal periodicity of the waves I will show diagnostics as the meridional and temporal average fields defined by
\begin{equation}
    \langle F\rangle = \int_0^{\frac{2\pi}{\omega}}\int_{-\frac{\pi}{2l}}^\frac{\pi}{2l} F \ dy\,dt. \label{average}
\end{equation}
\begin{figure}
    \centering
    \includegraphics[width=\linewidth]{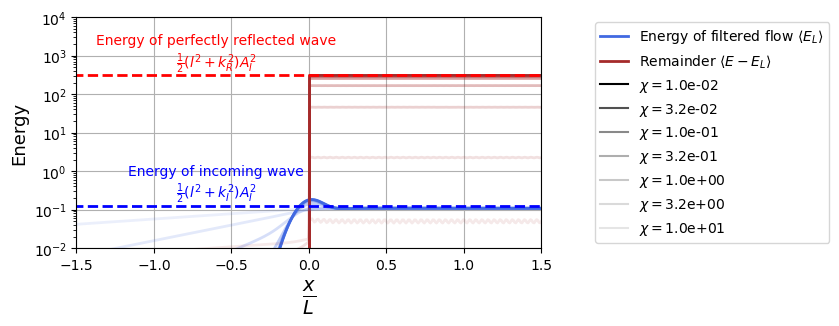}
    \caption{\textbf{Energy of filtered flow and energy of the remaining flow in the asymptotic limit $\mathbf{\chi \rightarrow 0}$.} The non-dimensional frequency has been set to $\omega = 0.02$ and wave energies are plotted for decreasing values of $\chi$. In the limit $\chi\rightarrow 0$, the energy of the porous wave ($x<0$) vanishes and the reflected wave grows ($0<x$) until saturating at an amplitude equal to that of the incoming wave.}
    \label{fig: energies}
\end{figure}
We use a generic, normalized gaussian filtering function
\begin{equation}
    G_L(x,y) = \frac{4\pi}{L^2}e^{-4\pi^2\frac{x^2 + y^2}{L^2}}.
\end{equation}
with a filter scale equal to
\begin{equation}
    L = \frac{1}{2}\left(\frac{2\pi}{\sqrt{k_I^2 + l^2}} +\frac{2\pi}{\sqrt{k_R^2 + l^2}}\right).
\end{equation}
Figure \ref{fig: energies} shows that applying this Gaussian filter to the wave field filters out the reflected wave but conserves the energy of the incoming wave (blue curve at $0 < x$). As $\chi$ decreases, the energy of the porous wave decreases (blue curve at $x<0$) and the energy of the reflected wave increases (red curve at $0 < x$). In the limit $\chi\rightarrow0$ the energy of the reflected wave saturates at the energy expected for a perfect reflection off a solid boundary, which corresponds to equal amplitudes between the incoming and the reflected waves.\\

\begin{figure}
    \centering
    \includegraphics[width=\linewidth]{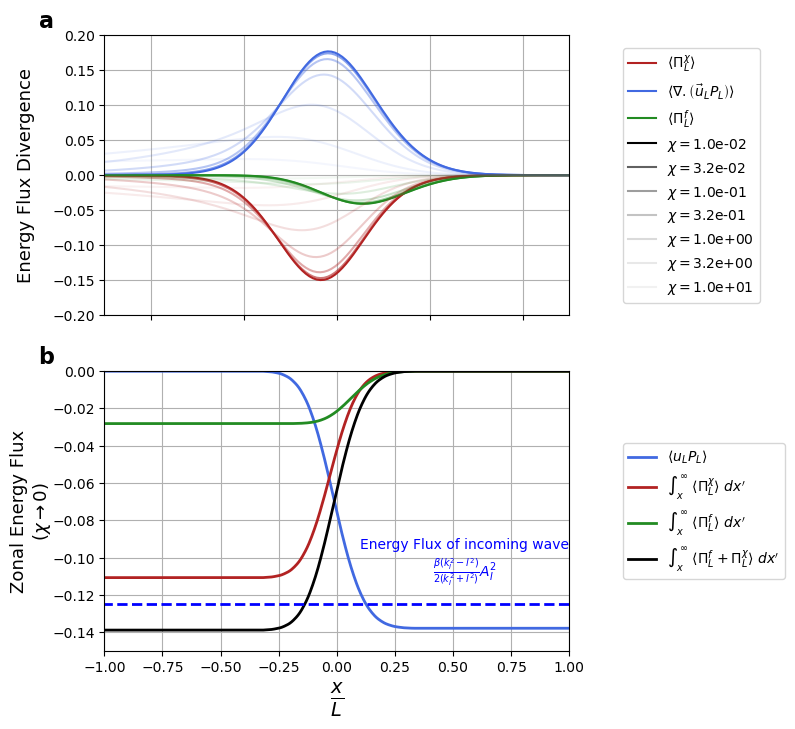}
    \caption{\textbf{Interscale energy budget of an incoming Rossby wave reflected off a boundary.} The local tendencies for the energy of the filtered flow from equation \eqref{Filtered_Energy} are shown in \textbf{a} for vanishing values of $\chi$. The main finding presented here is that the subgrid boundary flux $\Pi_{L}^\chi$ (red curve) converges in the limit of a solid boundary ($\chi\rightarrow 0$), implying that it is a well-defined physical quantity whose value does not depend on the exact value of the auxiliary parameter $\chi$. The zonally integrated budget of equation \eqref{cumulative_conservation} is shown in \textbf{b} only for the limiting case $\chi \rightarrow 0$, and illustrates that the downscale fluxes entirely evacuate the energy fluxed westward by the incoming wave. The slight mismatch between the filtered pressure flux and the energy flux of the incoming wave predicted by linear theory is thought to be due to finite-size effects and is expected to disappear at larger scale separation.}
    \label{fig: energetics}
\end{figure}

The main finding of this analysis lies in the explicit calculation of all the terms in the energy tendency equation \eqref{Filtered_Energy}, which are shown in figure \ref{fig: energetics}a. A priori it may not seem clear how the subgrid boundary flux $\Pi_L^\chi$ behaves in the asymptotic limit $\chi \rightarrow 0$, as it could diverge due to the appearance of $\chi$ in the denominator of the porous stress in \eqref{NS_porous}. However, figure \ref{fig: energetics}a shows that the flow $\vec{u}$ outside of the domain vanishes accordingly for $\Pi_L^\chi$ to converge to a finite curve as $\chi$ vanishes. This illustrates that the auxiliary parameter $\chi$ does not determine the value of $\Pi_L^\chi$, which confirms that $\Pi_L^\chi$ is a meaningful physical quantity. Indeed $\Pi_L^\chi$ balances the divergence of the Pressure flux of the filtered flow when approaching the boundary, illustrating that the energy of the incoming wave is transferred downscale to the reflected wave. The subgrid coriolis flux $\Pi_L^f$ only contributes negligibly to this downscale transfer, and is expected to vanish entirely at larger scale separation.\\

Another helpful way to understand the energy budget of the filtered flow is to look at the cumulative integral of the energy fluxes along the zonal direction. After applying the average in equation \eqref{average} to the energy budget in equation \eqref{Filtered_Energy} the meridional dependence drops out, and I can rewrite the budget as 
\begin{equation}
    \partial_x\left(\langle u_{L}P_{L}\rangle + \int_x^\infty \langle \Pi_{L}^\chi(x') +\Pi_{L}^f(x') \rangle\, dx'\right) = 0,
\end{equation}
implying that 
\begin{equation}
    \langle u_{L}P_{L}\rangle + \int_x^\infty \langle \Pi_{L}^\chi(x') + \Pi_{L}^f(x') \rangle\, dx' = C \label{cumulative_conservation}
\end{equation}
where $C$ is a constant. This means that the cumulative integral of the subgrid fluxes has to account for losses in the zonal pressure flux of the filtered flow. Figure \ref{fig: energetics}b shows that this balance holds and that the energy which is fluxed towards the west by the incoming wave is entirely transferred downscale by the subgrid fluxes close to the boundary. The majority of this downscale flux is mediated by $\Pi_{L}^\chi$.

\section{Conclusion}

In this document I showed that it is possible to derive closed scale-dependent energy budgets for bounded flows by combining the methodologies of volume penalization and flow filtering. This allows for the definition of a novel quantity, the subgrid boundary flux $\Pi_L^\chi$, which measures the interscale energy transfer occurring at boundaries. I also provided a proof-of-concept by applying this methodology to the reflection of Rossby waves, which revealed that $\Pi_L^\chi$ is a crucial term in the interscale energy budget of bounded flows that has thus far completely defied quantification.\\

The analytical simplicity of Rossby reflections permitted an explicit calculation of $\Pi_L^\chi$. I thus could show that $\Pi_L^\chi$ converges to a finite value in the limit $\chi\rightarrow 0$, which confirms that it is a well-defined physical quantity that does not depend on the value of the auxiliary parameter $\chi$. In more complex settings an explicit calculation may not be feasible. However, $\Pi_L^\chi$ can be implicitly deduced by extending the solution obtained close to a solid boundary with $\vec{u} = 0$ outside of the flow domain. All the non-singular terms of the filtered energy equation \eqref{Filtered_Energy} could then be calculated and $\Pi_L^\chi$ deduced implicitly as the remainder. The methodology provided here can therefore readily be extended to diagnose $\Pi_L^\chi$ in more complex numerical simulations or reanalyses products. The characterization of $\Pi_L^\chi$ in these cases will allow the design of an entirely novel class of subgrid-scale parametrizations.\\

An interesting avenue for further research is to study the behavior of $\Pi_L^\chi$ in oceanic boundary layers. A direct calculation of $\Pi_L^\chi$, possible for laminar models of oceanic boundary currents, can be a novel means to understand the dissipative anomaly present in the wind-driven ocean circulation \citep{miller2024gyre}. Furthermore, the application of coarse-graining to non-linear bounded flows would permit a concomitant quantification of both the linear and non-linear mechanisms of interscale energy transfer. This would permit to revisit the long-standing question of how inverse cascades in the ocean interior \citep{scott2005direct} compete with down-scale cascades occurring close to continental boundaries \citep{evans2022dissipation}.
Last, the methodology outlined here also permits a systematic study of the appropriate boundary conditions for turbulent large-eddy simulations \citep{hausmann2025formally}, which remains an open problem in oceanic circulation models \citep{chassignet2008gulf}. \\

\newpage
\bibliography{bibio}

\begin{thebibliography}{10}

\bibitem{scott2005direct}
Robert~B Scott and Faming Wang.
\newblock Direct evidence of an oceanic inverse kinetic energy cascade from satellite altimetry.
\newblock {\em Journal of Physical Oceanography}, 35(9):1650--1666, 2005.

\bibitem{sagaut2006large}
Pierre Sagaut.
\newblock {\em Large eddy simulation for incompressible flows: an introduction}.
\newblock Springer, 2006.

\bibitem{stommel1948westward}
Henry Stommel.
\newblock The westward intensification of wind-driven ocean currents.
\newblock {\em Eos, Transactions American Geophysical Union}, 29(2):202--206, 1948.

\bibitem{pedlosky2013geophysical}
Joseph Pedlosky.
\newblock {\em Geophysical fluid dynamics}.
\newblock Springer Science \& Business Media, 2013.

\bibitem{aluie2018mapping}
Hussein Aluie, Matthew Hecht, and Geoffrey~K Vallis.
\newblock Mapping the energy cascade in the north atlantic ocean: The coarse-graining approach.
\newblock {\em Journal of Physical Oceanography}, 48(2):225--244, 2018.

\bibitem{hausmann2025formally}
Max Hausmann and Berend Van~Wachem.
\newblock A formally exact wall-boundary condition in large eddy simulations using volume filtering.
\newblock {\em Journal of Fluid Mechanics}, 1022:R4, 2025.

\bibitem{engels2015numerical}
Thomas Engels, Dmitry Kolomenskiy, Kai Schneider, and J{\"o}rn Sesterhenn.
\newblock Numerical simulation of fluid--structure interaction with the volume penalization method.
\newblock {\em Journal of Computational Physics}, 281:96--115, 2015.

\bibitem{schneider2005decaying}
Kai Schneider and Marie Farge.
\newblock Decaying two-dimensional turbulence in a circular container.
\newblock {\em Physical review letters}, 95(24):244502, 2005.

\bibitem{miller2024gyre}
Lennard Miller, Bruno Deremble, and Antoine Venaille.
\newblock Gyre turbulence: Anomalous dissipation in a two-dimensional ocean model.
\newblock {\em Physical Review Fluids}, 9(5):L051801, 2024.

\bibitem{evans2022dissipation}
D~Gwyn Evans, Eleanor Frajka-Williams, and Alberto~C Naveira~Garabato.
\newblock Dissipation of mesoscale eddies at a western boundary via a direct energy cascade.
\newblock {\em Scientific Reports}, 12(1):887, 2022.

\bibitem{chassignet2008gulf}
E~Chassignet and D~Marshall.
\newblock Gulf stream separation in numerical ocean models.
\newblock {\em Geophysical Monograph Series}, 177, 2008.

\end{thebibliography}
\newpage

\section{Appendix: Analytical Solution to Rossby Wave Reflection off a Porosity Step}
\label{appendix}
The linear flow model in \eqref{NS_porous} permits an analytical treatment of the behaviour of Rossby waves reflecting off a porosity step. We follow the methodology where an incoming wave is imposed and additional waves with the same temporal frequency $\omega$ and meridional wavenumbers $l$ are sought such that the total flow satisfies the boundary conditions.\\

 A first simplification of \eqref{NS_porous} consists in introducing $\psi$ as the stream function such that $(u,v) = (-\partial_y\psi, \partial_x\psi)$. Taking the curl of \eqref{NS_porous} inside the flow domain ($0 < x$) then yields the vorticity equation
\begin{equation}
    \partial_t\nabla^2\psi + \beta\partial_x\psi = 0. \label{beta-plane_dimensional}
\end{equation}
I will investigate wave-like solutions of the form
\begin{align}
    \psi = \mathcal{R}\left(Ae^{i(k x + l y -\omega t)}\right).\label{waves}
\end{align}
Here, $k$ and $l$ are the zonal and meridional wavenumbers of the wave, $\omega$ its the temporal frequency and $A$ its amplitude. The real part of the exponential expression is denoted by $\mathcal{R}$. Inserting \eqref{waves} into \eqref{beta-plane_dimensional} I obtain the dispersion relation of Rossby waves,
\begin{equation}
    \omega = -\frac{\beta k}{k^2 + 1}.\label{Rossby_dispersion_dimensional}
\end{equation}
which is shown in figure \ref{fig:Rossby_disp}. For a given frequency $\omega$ there are two values of $k$ that satisfy the dispersion relation. There is a long wave solution $k_I$,
\begin{equation}
    k_{I} = \frac{-\beta - \sqrt{\beta^2 - 4l^2\omega^2}}{2\omega}.
\end{equation}
And a short wave solution $k_R$,
\begin{equation}
    k_{R} = \frac{-\beta + \sqrt{\beta^2 - 4l^2\omega^2}}{2\omega}. 
\end{equation}
These waves transport energy into the direction of the group velocity, whose zonal component is $c^g_x = \partial_k\omega$. As visible in figure \ref{fig:Rossby_disp}, the long wave has a $k_I$ in the range where $\omega$ is decreasing as a function of $k$ and thus transports energy to the west. Vice versa, the short wave has a $k_R$ in the range where $\omega$ is increasing as a function of $k$ and thus transports energy to the east. The typical energetics of Rossby wave reflection at solid western boundaries are thus the following: A long, incoming wave with wavenumber $k_I$ transports energy to the western boundary, where it is transferred downscale and transported eastward by a short, reflected wave with wavenumber $k_R$.\\

Here I will investigate the reflection off a step in porosity. Taking the curl of \eqref{NS_porous} outside of the domain ($x < 0$) yields the porous vorticity equation
\begin{equation}
    \partial_t\nabla^2\psi + \beta\partial_x\psi = -\frac{1}{\chi}\nabla^2\psi. 
\end{equation}
We can again look for solutions of the form \eqref{waves} and obtain the porous dispersion relation 
\begin{equation}
    \omega = -\frac{\beta k}{k^2 + l^2} - i\frac{1}{\chi}. \label{Rossby_dispersion_porous}
\end{equation}
This modified dispersion relation admits the complex, decaying wavenumber given by
\begin{equation}
    k_P = -\frac{\beta - \sqrt{\beta^2 - 4l^2(\omega + i/\chi)^2}}{2(\omega + i/\chi)} \label{total solution}
\end{equation}
Here I chose the solution that decays towards the west as I am interested in reflections at western boundaries. This third branch of the dispersion relation is also shown in figure \ref{fig:Rossby_disp}, and suggests that outside of the domain a third, porous wave with wavenumber $k_P$ is excited when a Rossby wave impinges on a step in porosity.\\

\begin{figure}[htb!]
    \centering
    \includegraphics[width=\linewidth]{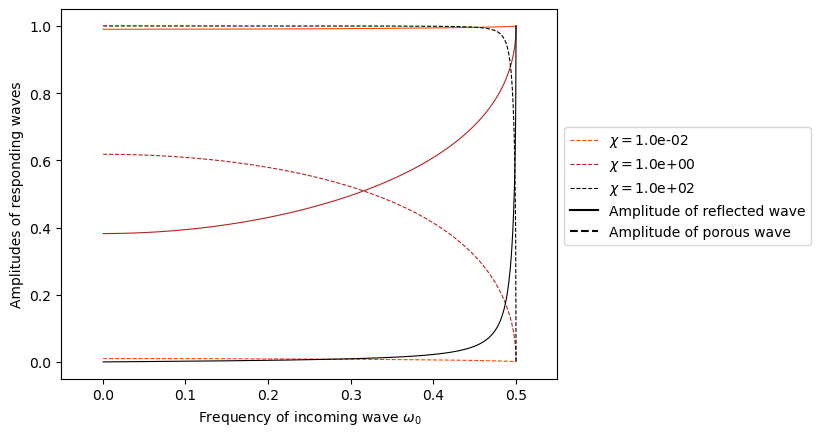}
    \caption{\textbf{Amplitudes of reflected and porous waves for three exemplary values of the porosity $\mathbf{\chi}$}. The meridional wavenumber $l$ and vorticity gradient $\beta$ have been set to unity. The energy of the incoming wave is reflected for high frequencies and transmitted for low frequencies, but the cutoff moves to lower frequencies as $\chi$ decreases until eventually all frequencies are reflected.}
    \label{fig: frequ_response}
\end{figure}

The wave solution I seek is thus expressed analytically as
\begin{align}
    \psi^+ = \mathbf{R}\left(A_Ie^{i(k_Ix + l y -\omega t} + A_Re^{i(k_Rx + ly - \omega t}\right) \ &\text{ inside the domain ($0 < x$) and }\\
    \psi^- = \mathbf{R}\left(A_Pe^{i(k_Px + l y - \omega t}\right) \ &\text{ in the porous medium ($x< 0$).}
\end{align}
We now aim to determine the complex amplitudes of reflected ($A_R$) and porous waves ($A_P$) as a function of the amplitude of the incoming wave ($A_I$). Their expressions are obtained by imposing boundary conditions at the interface at $x=0$. First, I impose continuity of the streamfunction to obtain a solution with finite energy,
\begin{equation}
    \psi^+(x = 0)\  = \psi^-(x = 0).
\end{equation}
The second boundary condition follows from the conservation of circulation around an infinitesimal loop across the porosity step, which yields
\begin{equation}
    \partial_tv^+(x = 0) -\partial_tv^-(x = 0) = \frac{v^-(x=0)}{\chi}.
\end{equation}
Together, these two conditions yield the following expressions for the wave amplitudes
\begin{align}
    A_R &= \frac{\omega\chi(k_I - k_P) - ik_P}{\omega\chi(k_P - k_R) + ik_P}\, A_I, \\
    A_P &= \frac{\omega\chi(k_I - k_R)}{\omega\chi(k_P - k_R) + ik_P}\, A_I
\end{align}
The limit of a solid boundary is achieved when $\chi = 0$, which corresponds to a perfect wave reflection with $A_P = 0$ and $A_R = A_I$. At $\chi \rightarrow \infty$, the wave is completely transmitted instead. Figure \ref{fig: frequ_response} illustrates that at finite values of $\chi$ the porosity step reflects waves of higher frequency more strongly, and thus behaves as an effective low-pass filter for the incoming waves.\\

For the explicit tracing of all terms in the energy equation \eqref{Filtered_Energy} I also require an explicit expression of the pressure $P$. Rescaling and taking the divergence of \eqref{NS_porous}, I obtain that $P$ governed by 
\begin{equation}
    \nabla.\left(\vec{f}\times\vec{u}\right) = \nabla^2P.
\end{equation}
After inserting the streamfunction I integrate the left hand side by parts to obtain
\begin{equation}
    \nabla^2\left((f_0 + \beta y)\psi\right) - \beta \partial_y\psi = \nabla^2P.
\end{equation}
We can thus define the compensated pressure $P' = P - (f_0 + \beta y)\psi$. It satisfies
\begin{equation}
    - \beta \partial_y\psi = \nabla^2P'.
\end{equation}
As all of the waves in our solution are expressed as $\psi = Ae^{i(kx+ly-\omega t)}$ I look for similar particular solutions of the shape $P' = A^Pe^{i(kx+ly-\omega t)}$. Inserting this expression finally yields
\begin{equation}
    \frac{i\beta l}{k^2 + l^2} A = A^P.
\end{equation}
We thus obtain the particular solutions for the pressure of the incoming, reflected and porous waves as
    \begin{align}
    P_I &= \left(\frac{i\beta l}{k_I^2 + l^2} + f_0 + \beta y\right)A_Ie^{i(k_Ix + ly - \omega t)}\\
    P_R &= \left(\frac{i\beta l}{k_R^2 + l^2} + f_0 + \beta y\right)A_Re^{i(k_Rx + ly - \omega t)}\\
    P_P &= \left(\frac{i\beta l}{k_P^2 + l^2} + f_0 + \beta y\right)A_Pe^{i(k_Px + ly - \omega t)}.\\
\end{align}
The pressure has to satisfy two boundary conditions. First, $P$ has to be continuous at $x=0$,
\begin{equation}
    P^+(x = 0) = P^-(x = 0). \label{Pressure_continuity}
\end{equation}
The second boundary condition is obtained by subtracting the zonal component of the Navier-Stokes equation at the right hand side of the interface from the expression at the left hand side. Requiring continuity of the normal velocity component $u$ (conservation of mass), one thus obtains
\begin{equation}
    -(f_0+\beta y)\left(v^+(x = 0)-v^-(x = 0)\right) = -\partial_x P^+(x = 0) + \partial_x P^-(x = 0) + \frac{1}{\chi}u^-(x = 0). \label{NS_continuity}
\end{equation}
To lift these two conditions, I add two zero-divergence terms to the pressure.
\begin{align}
    P^+ &= P_I + P_R + B^+e^{-lx + i(ly - \omega t)}\\
    P^- &= P_P + B^-e^{lx + i(ly - \omega t)}.
\end{align}
Solving \eqref{Pressure_continuity} and \eqref{NS_continuity} ultimately yields
\begin{align}
    B^+ &= \frac{1}{2}\left(\frac{i A_P}{\chi} + f_0(A_P - A_I - A_R) + \beta\left(\frac{A_P}{k_P - i l} - \frac{A_I}{k_I - i l} - \frac{A_R}{k_R - i l}\right)\right)\\
    B^+ &= \frac{1}{2}\left(\frac{i A_P}{\chi} - f_0(A_P - A_I - A_R) + \beta\left(\frac{A_P}{k_P + i l} - \frac{A_I}{k_I + i l} - \frac{A_R}{k_R + i l}\right)\right)
\end{align}
This completely determines the solution of a wave reflection off a discontinuity in porosity.\\

The Coriolis parameter is in principle a free parameter of the problem, but doesn't impact the pressure work in equation \eqref{Filtered_Energy} as it only determines the geostrophic pressure field. For the results presented in the main text it is thus set to the arbitrary value $f_0 = 1$. I also set $\beta = 1$, which can be seen as a means to non-dimensionalize the problem. The solution analyzed in the main text and depicted in figure \ref{psi_illustrate} is constructed as the superposition of two meridional wave numbers $l = \pm 1$ in order to avoid meridional energy fluxes.\\

\end{document}